\def\fenome{phenomenal entity}
\def\fenomes{phenomenal entities}

\documentclass{article}
\usepackage[hypertex]{hyperref}
\title{Qualia are Quantum Leaps}
\author{George Svetlichny\footnote{Departamento de Matem\'atica, Pontif\'{\i}cia Universidade Cat\'olica, Rio de Janeiro, Brazil \newline
svetlich@mat.puc-rio.br \hfill \url{http://www.mat.puc-rio.br/\~svetlich}}}
\begin{document}
\maketitle

\begin{abstract}
I contemplate the idea that the subjective world and  quantum state reductions are one and the same. If true, this resolves with one stroke both the quantum mechanical measurement problem and the hard problem of consciousness.
\end{abstract}
\vskip 12pt
{\obeylines
\hfill{\it There is a crack, a crack in everything}
\hfill{\it That's where the light gets in}
\hfill Leonard Cohen
}

\vskip12pt
Two scientific problems have been with us for a very long time with no apparent resolution. One is the measurement problem of quantum mechanics whereby the physical evolution of a system is subject to  apparent discontinuities while, presumably, undergoing a ``measurement" or ``observation". No theoretical or empirical investigation has thrown any convincing light on the nature of this discontinuity, nor whether it actually exists.  The other is what is known as the ``hard problem of consciousness": how to explain how we, as physical beings,  can have phenomenal sensation of enjoying a flower's perfume and color, loving a mate, and hating Mondays. It is generally deemed that these sensations are not  attributes of  physical objects but that of our consciousness. They are contents of our mind.

Both discontinuities and sensations are beset with conceptual and scientific difficulties whose attempt at resolution up to now have not produced any insight, have not provided anyone with an ``aha" experience later shared by many. What if both problems were the same? This would mean that there truly are discontinuities and that they are sensations.

{\it Caveat lector:\/} This is a radical idea and at first blush not very believable. However, it has a strong seductive quality and a certain superficial air of plausibility. The appeal and  hope is that two apparently insoluble problems may solve each other. If there is any truth to what's written below, it is probably contained in only a small part of what follows. Our subjective world is centered around our bodies, which, along with the subjective content are the results of millions of years of evolution producing highly complex structures. There is no proper vocabulary to apply to join the two along the lines suggested and I am forced to fall back upon imprecise jargon borrowed from various fields. Some of the text has to be considered  quasi-metaphorical, and some {\em is\/} unabashedly truly metaphorical. Even so some remarkable things do emerge from this idea.

There is no single term that refers to  entities of the phenomenal world, something that would distinguish them from physical objects such as molecules, people and elephants. Philosophers have come to denote  subjective sensations as {\em qualia\/}. It comes with too much baggage and I will not use it.
 I've used it in the title to be catchy and it does state correctly what is proposed here, but there is much more. Not wanting to coin a term nor use acronyms I'll just settle for {\em phenomenal entity\/} until something better comes along.

As I look at my bookshelf I see a book and I am aware of two things about it, its position on the shelf and its green color. As far as the position is concerned, this is deemed to be a physical attribute. The book resides at a particular site in space in relation to the walls of my office. The color green is deemed a pure sensation, not a physical attribute at all but a \fenome. However, the book's position should not be confused with my sensation of its position, so obviously both sensations are \fenomes. While much has been written about the non-physical phenomenal quality of color and odors, sensations of position and velocity are systematically left out, a curious bias.  From a unifying posture, either there is less physicality  corresponding to the sensed position, or more to that of green. A little of both is probably the truth, but the latter position which asserts rather than denies seems potentially more fruitful and I am led to propose that in fact the \fenome\  green is physical. It doesn't correspond to a physical object nor to an attribute of one but to a quantum mechanical discontinuity. The \fenome\  of position also corresponds to a discontinuity but a different one. I, as a subject, am also a discontinuity, that's why I'm aware of the other two. All three entities are the same kind of thing.

These last sentences need more elaboration.

Firstly, the usual deterministic unitary evolution of quantum states entangles all parts.  One must think of the quantum state involved as the state of the universe, so the discontinuities that I speak of must be considered discontinuities of the universe, and it is so that I'll treat them.

Secondly, I am not resurrecting the proposal that consciousness causes state collapse, nor what may {\em seem\/} to be my proposal that state collapse gives rise to consciousness as a separate entity. I am identifying the two: consciousness {\em is\/} state collapse. We, as subjects, abide in the discontinuities of the universe, or better put, {\em are\/}  discontinuities of the universe, we do not dwell in a universe, we dwell {\em between\/} universes, or more to the point, we {\em are\/} the  ``between-universes''.

To say that state reduction {\em causes\/} a sensation, or that it {\em gives rise\/} to consciousness is to push the problem along to another level. We are forced to ask: {\em Who\/} is having the sensation and what is it? {\em Who\/} is conscious and what is consciousness? Same old problem again. To resolve it one must stop this eternal withdrawal, cut to the chase,  and just say the word {\em is\/} at some point. I am saying it here. The buck stops at quantum state reduction. Let's see what results from this hard position.

The discontinuities that are the \fenomes\  associated to the perceived book of course do not take place in the book. All my sensations seemingly arise  in my body, and most would say  my brain. It's  thus natural to think that the quantum discontinuities that are the \fenomes\  in my consciousness are ones involving the quantum state of my body and especially my brain. The green color of the book would be due to my retina and my brain and undoubtedly much other neurological paraphernalia.  Molecules in my body are thus the immediate objects of my consciousness. It is these that I, as a subject, touch intimately. It is the state reductions of these that are my sensations. Just how it is that I see the book ``out there" is a separate problem. The  objective physical world is covered over by a field of \fenomes\  that appear to be properties of things ``out there'' and placed there as a kind of skin. In truth the ``things out there''  do not posses the attributes that we naively say they do. They have no position, no color, no odor and produce no sound. This compelling illusion has obvious survival value and so must be the result of the evolutionary development of my body,  my brain and my mind.

I do not deny the existence of an objective physical world nor the existence of state collapse outside my body. There is collapse occurring in rocks, swamp gas, and cherry blossoms. I am not saying that these things feel pain, enjoy the breeze or contemplate the beauty of prime numbers. They have {\em the same kind\/} of stuff, state reductions, that make up our minds which do feel, enjoy and contemplate, just as they have {\em the same kind\/} of stuff, atoms, that make up our bodies. The difference is in the organization of the two kinds of things. We as humans are not privileged sole owners of a transcendental stuff that dreams are made of, we are {\em gatherers and hoarder\/} of the kind of stuff that dreams are made of and which is spread around everywhere across the galaxies, just as we are {\em gatherers and hoarders\/} of the kind of stuff our bodies are made of, which is also spread around everywhere across the galaxies. There is no sharp division between living and non-living matter, there's a continuum, and there is likewise no sharp division between matter that has subjective content and that which hasn't. Just as I don't consider little clumps of molecules on the floor or in raindrops as organisms, I shouldn't consider the state reductions in them as minds. They are just ingredients from which organisms and minds are made of.

The brain is now imagined as the central processor of a state-reduction system, roughly the body. It allows certain collapses to occur and not others, and generally controls and coordinates them. More importantly it allows for a special \fenome, the sensed self, to reside. It is a product of evolution and the system of collapses it coordinates is as complex, if not more so, as the rest of the body. If the brain is a quantum mechanical processor, it needs to create precisely tailored  quantum states to sustain the phenomenal world through reductions. One sees constant activity in the brain, signals, electrical and chemical, coursing through it incessantly. One must now think of many of these as the classical communication part of a sophisticated quantum processing system. The underlying quantum part, undoubtedly involving highly entangled states, has not yet been identified. It operates at body temperature and so Nature has found ways to resolve the problems of decoherence and fragility of quantum states. Or else, there are aspects of quantum mechanics, especially of state collapse that we've not yet discovered. The body's underlying quantum state needs to be refreshed as there are constant reductions, which are the \fenomes. Presumably this happens during default activity and sleep.

Given the enormous complexity of both the body and the phenomenal world it harbors and this radical proposal of their relationship, I cannot hope to say much that is correct and meaningful in such an incipient state of contemplation. Some things can be said nevertheless and I do so in the paragraphs that follow. They will be somewhat sketchy and disconnected. I've avoided much elaborations, as this would undoubtedly lead me far astray. As a further warning I must say I'm taking a minimalist and extreme position in this essay. This position is that all \fenomes\  are state reductions and all state reductions are \fenomes. What can be immediately inferred from this is astonishing and often seemingly preposterous. The truth, if there be any, undoubtedly lies on this side of preposterousness and hopefully this way some of it can be caught.

The other-worldly quality of \fenomes\  becomes understandable. Collapse itself is other-worldly, it is outside the dynamical laws of unitary quantum mechanics. It is subject to some sort of ontological indeterminism. One can only say ``collapse happens" but what provokes it and when, is the central mystery. Some probabilistic regularities are seen but that's about all we know.  It is the ``between-universes". One cannot join an other-worldly realm to a worldly one except as some sort of ephemeral epiphenomenon, otherwise there would be no need for the modifier ``other". Two other-worldly realms however could truly join without causing conceptual perplexities and paradoxes. True the joined realms are still other-worldly, but now they are within the province of a still  much unexplored physics and with time can become as worldly as the worldly realm is now. Physics then truly becomes a science of everything.

One understands the high dimensionality of the field of \fenomes. To code all the information present would require untold billions of bits. One does not perceive this as a vector in a hugely dimensional space because much of the field is painted over the external world to create the illusion that the \fenomes\  reside ``out there" and so we naively think we are not rendering but sensing. This high dimensionality is in keeping with the exponential growth of the dimension of the Hilbert space describing the internal degrees of freedom of multipartite systems, and with the infinite dimensional Hilbert spaces describing the spacial degrees of freedom. In other words, there's lots of space in Hilbert space.

The non-local distributed nature of cognition becomes understandable. As I look at my bookshelf I see a high resolution image of that part of the visual field that falls on the fovea of my retina. Sub-millimeter detail is visible from four feet away. Yet there is no such fine geometrical placements at ends of the nerves that lead to the visual cortex. If one sees with the brain how is this possible? The truth is that one sees with the whole distributed network, and not with any part of it. This is in keeping with the nonlocal nature of entangled quantum states. I have been careful not to separate too much the body and the brain. It is the whole body, and possibly some things external to it, that is the quantum processing system. The brain plays only a part, important as that part may be. The mind-body relation now has a new form. The mind {\em is\/} the field of \fenomes\  which {\em are\/} the coordinated quantum state reductions in the body.

Some phenomenal entities are reflexive, that is self-referential. I am conscious of consciousness, I think  of thoughts, I exert my will on my will, as when I metaphorically  bite my tongue. Less clearly one loves love, hates hate and fears fear. But one can't green green. What allows for reflexive entities is not clear. Reflexivity seems to be a sign of universality. One can be conscious of anything, hence of consciousness itself. One can think of anything hence of thought itself. One can fear anything, hence fear fear itself. This universality also explains how one can hate love or love pain. Are there, even if to minuscule degrees, universal \fenomes\  present in all collapse instances?  Space-time is a universal presence for all systems, and its degrees of freedom participate in all collapse instances. Space-time is a reflexive entity, a container that contains itself. Gravity, a manifestation of space-time,  is a force that acts on itself. The reflexive quality of some \fenomes\  may have its roots here.

There is a certain current fashion in thinking that consciousness and will are emergent entities, that they only arise once one has a system of sufficient complexity. That is, consciousness and will are epiphenomena, supervenient on the physical world. This of course is antithesis to the present proposal, which, as a natural corollary, would claim that both consciousness and will and other reflexive \fenomes\  are present, to some degree, from the very start. As a consequence, evolution can start anywhere with the first state collapse. Once a system locks onto  ontological indeterminism, the hallmark of quantum state-reduction, it has reproductive advantage as it acquires a wider repertoire of now unpredictable behavior. It acquires a will (more on this later). Thus life should arise anywhere that the physical conditions allow for its existence. The same consideration apply to intelligent life for once life harbors state reductions in an orderly way, evolution takes it to brains and awareness and we humans is one of the results. Life becomes a ``measuring instrument'' of the universe. This gives it an advantage as actualized choice from a potentiality of possibilities makes a difference.

Some comments on the previous paragraph is necessary. One should not think that the state reduction of say a single photon is conscious and has a will the same way we do. Just as our bodies are made up of a huge number of molecules one must think that our subjective world is also a construct with a huge number of constituent that we're not even aware of. One of these constructs is the consciousness that shines its light on all things perceived when we're awake, and another is what we call the will which acts on possibilities presented to us by the ontological indeterminacy that characterizes state reduction. What the collapse of a single photon has are the elementary constituents of what are the constructs in {\em our\/} phenomenal worlds.

Explanation of consciousness based merely on complexity of networks or software are futile. This would be saying that by reproducing the classical communication or processing part of a quantum processing system one could reproduce all the workings of the quantum part also. Conscious robots that only have sophisticated software running on classical processing systems are preposterous.

The sensation of green must be the same in all people that have the usual trichromatic visual system and homologous brain wiring. Since they have the same molecules, arranged in the same way, essentially the same brain construction and same classical communication system, they have the same state collapse when light of the same spectral quality and from the same scene enters the eyes, and so they have the same same sensation. In principle this can be verified experimentally. If two brains can be induced to share enough entangled pairs of molecules, through entanglement swapping say,  and an appropriate classical communication channel set up, then \fenomes\  can be quantum teleported from one to the other and one would see through the eyes of the other. This is not feasible with present day technology, but in time it can be achieved. One could also in principle teleport \fenomes\  from animals, plants, and even inanimate objects, turning us into intimate witnesses of the inner workings of things. One truly should be able to know a little of what it's like to be a bat.

Sophisticated quantum computers running at room temperature are possible, the brain is proof and prototype. It seems very much like a one-way quantum computer, that is, one based on measurements. It has to be such to a large extent, as it is precisely the measurements that maintain the phenomenal world. The highly entangled resource that such computers run on has to be replenished almost constantly. This, as was stated before, presumably is the task of the default mode and sleep.

Memory is a facsimile of past \fenomes. This implies that either there is approximate quantum cloning of quantum states, which are then stored as quantum memories, or there is a certain amount of quantum tomography after which enough classical information can be stored to be able to recreate a facsimile of the state. There are many types of memory so many strategies are undoubtedly used. In any case, decade-long approximate quantum memories at room temperature are possible as again the brain is the proof and prototype.

Artificial consciousness is possible. The development and birth of a human infant is the proof and prototype. The process of development builds, using ordinary matter scattered around on this planet,  a complex physical object containing a state-reduction processing system which makes possible a subject and its phenomenal world. Any other procedure that create a sufficiently complex state-reduction processing systems, using the same or similar ordinary matter, would create an artificial sentient being fully equipped with its own phenomenal world.

Will would be the \fenome\ that can act, but can it? Freedom of will is obviously problematic in a deterministic world, but it still can be problematic in a nondeterministic one also. State reductions are presumably governed by ontological indeterminism, but even so, are subject to the Born rule of quantum probabilities, a precise statement of outcome behavior. So maybe Nature does play dice  and the hapless subject is simply dictated by its toss deluding itself that it has acted. For free will to have a place one has to get rid of the dice. Probability is not truly a scientific discipline. Conformity to precise random distributions has no compelling argument. Ontological indeterminism need not correspond to precise probability distributions, and could be more like ``ontological capriciousness". The Born rule is empirically found to hold, but never tested under such complex situations as state-reduction in the human brain. If we as subjects do have free will it will manifest itself as departures from the Born rule. We would be actively choosing from among the possible states that could result from a quantum state reduction in our body, and not just subjecting ourselves to an external toss of the dice. Nature would not be tossing dice but providing us with opportunities. In principle this is an empirical question.

The body is a unit in spite of constant renewal of matter. The unity is maintained basically by electromagnetic forces between molecules. The mind is a unit in spite of being a constantly updated process. It's not clear what maintains this unity though the body's unity goes a long way in explaining it. Being quantum state reductions in the body, something about the reduction process is responsible for maintaining integrity. This is likely a part of quantum physics that we've not yet learned about. It may be that here too a departure from the Born rule occurs, the mind, so to speak, wills itself stay cohesive.

To sum up, the phenomenal world is identified with quantum state reductions. Our bodies and especially our brains are quantum collapse processors that create us as phenomenal subjects existing in a phenomenal word, discontinuities in the universe. The study of the phenomenal world gets unified with that of the physical world in a unique science more akin to physics than to anything else. Empirical access to the phenomenal world is possible under the same terms as to the physical world.

\section*{No References}
The ideas presented here are radical. Either what is written above resonates to some degree with the thoughts and conceptions of the reader on its own terms, or else, all is an idle daydream. Any reference would be a needless distraction.

\section*{Acknowledgements}This research received partial financial support from the Funda\c{c}\~ao de Amparo \`a Pesquisa do Estado do Rio de Janeiro  (FAPERJ).

\end{document}